\documentstyle[twocolumn,prl,aps,epsfig]{revtex}

\begin{document}

\draft
\preprint{}

\twocolumn[\hsize\textwidth\columnwidth\hsize\csname
@twocolumnfalse\endcsname

\title{Superfast front propagation in reactive systems 
	with anomalous diffusion}
\author{Rosaria Mancinelli$^{1,2}$, Davide Vergni$^{2,3}$
 and Angelo Vulpiani$^{1,2}$
}
\address{
  $^1$ Dipartimento di Fisica,
Universit\`a di Roma `La Sapienza',  P.le A. Moro 2, I-00185, Roma,
Italy.\\
  $^2$ Istituto Nazionale Fisica della Materia UdR and SMC Roma 
`La Sapienza'\\
  $^3$ Istituto Applicazioni del Calcolo (IAC) - CNR, 
Viale del Policlinico 137, I-00161, Roma,
Italy.\\
}

\date{\today}
\maketitle
\begin{abstract}
We study a reaction diffusion system where we consider a non-gaussian
process instead of a standard diffusion. 
If the process increments follow a probability distribution 
with tails approaching to zero faster than a power law, the
usual qualitative behaviours of the standard reaction diffusion
system, i.e., exponential tails for the reacting field 
and a constant front speed, are recovered. 
On the contrary if the process has power law tails,
also the reacting field shows power law tail 
and the front speed increases exponentially with time.
The comparison with other reaction-transport systems which exhibit
anomalous diffusion shows that, not only the presence of anomalous
diffusion, but also the detailed mechanism, is relevant for the front
propagation.
\end{abstract}

] \narrowtext 


The propagation of fronts generated by a reaction-transport system
has a considerable interest in a large number of chemical,
biological and physical systems~\cite{Murray}. One important 
model is the advection-reaction-diffusion equation (ARD)
\begin{equation}
\partial_t \theta  + ({\bf u} \cdot \nabla)
\theta = D_0 \nabla^2 \theta + f(\theta)/\tau ,
\label{eq:ard}
\end{equation}
where $D_0$ is the molecular diffusivity, ${\bf u}$ is a given
incompressible velocity field and the term $f(\theta)/\tau$ describes
the production process, whose typical time is $\tau$.  The
scalar field $\theta$ represents the fractional concentration of the
reaction's products: $\theta=1$ indicates the inert material,
$\theta=0$ the fresh one and $0 < \theta < 1$ means that fresh
materials coexist with products. Usually one considers $f(\theta)$
with an unstable fixed point at $\theta=0$ 
and a stable one at $\theta=1$~\cite{Xin,Peters}.
 
Eq.~(\ref{eq:ard}) was originally introduced (in the case 
${\bf u}={\bf 0}$) by Fisher and Kolmogorov, Petrovskii and Piskunov
(FKPP) with $f(\theta)=\theta(1-\theta)$~\cite{FKPP}.
Then, the ARD equation has been widely studied in the context
of combustion, population growth, aggregation and deposition.
\\
In absence of stirring $({\bf u}={\bf 0})$ it can be shown
that a front, replacing the unstable state by the stable one,
moves with a speed $v_0 = 2\sqrt{D_0 f'(0)/\tau}$. This result
is valid whenever $f(\theta)$ is a positive convex function
$(f''(\theta) < 0)$ with two fixed points. 
For a non-convex production term only an upper and lower bound 
for $v_0$ can be provided~\cite{Xin}.
\\
A more interesting situation occurs in the presence of stirring
$({\bf u}\neq{\bf 0})$: the front propagates with an average
limiting speed $v_f$ larger than $v_0$. 
Under general conditions~\cite{Majda,Castiglione}, 
the advection-diffusion equation, i.e., 
$\partial_t \theta  + ({\bf u} \cdot \nabla)\theta = D_0 \nabla^2 \theta$, 
has the same asymptotic feature of a diffusion equation. One has that
the field $\langle \theta \rangle$ (obtained with an average on
volumes of linear size much larger than the typical length of ${\bf
u}$) at large time obeys a Fick equation:
\begin{equation}
   \partial_t \langle \theta \rangle = 
   \sum_{i,j} D^e_{ij} \partial^2_{ij} \langle \theta \rangle\,\,.
   \label{eq:fick}
\end{equation}
The eddy diffusion coefficients, $D^e_{ij}$, contain all the
(often non-trivial) effects of the velocity field ${\bf u}$.
Eq.(\ref{eq:fick}) states that a 
test particle described by the Langevin equation:
\begin{equation}
   {{\mathrm d}{\bf x} / {\mathrm d}t} = 
   {\bf u} + \sqrt{2D_0} {\mbox{\boldmath{$\eta$}}}
   \label{eq:langevin}
\end{equation}
where {\boldmath{$\eta$}} is a white gaussian noise with $\langle
\eta_i \rangle = 0$ and $\langle \eta_i(t) \eta_j(t') \rangle =
\delta_{ij} \delta(t-t')$, follows, at large time, a Brownian motion
$\langle (x(t)-x(0))^2 \rangle \simeq 2 D_{11}^e t$, where we suppose
that the first coordinate is the propagation direction.  
As a consequence of the asymptotic diffusive behaviour it is possible to
show that the front propagates with a finite $v_f \leq 2
\sqrt{D_{11}^e f'(0)/\tau}$~\cite{Abel,Constantin}.
\\
If the conditions for the validity of standard diffusion do not hold,
anomalous diffusion can be observed,
i.e., 
\begin{equation}
 \langle (x(t)-x(0))^2 \rangle \sim t^{2\nu}
 \label{eq:anodif}
\end{equation}
with $\nu > 1/2$ (super-diffusion)~\cite{Castiglione,Zaslavsky}.
 There are at least three known mechanisms
leading to the anomalous diffusion~\cite{Majda,Castiglione,Zaslavsky,Levy,Metzler,standard,Gawedsky}:\\
  {\bf a)} an infinite variance of the velocity field ${\bf u}$;\\
  {\bf b)} an infinite memory, i.e., the velocity-velocity
            correlation function has a non integrable tail;\\
  {\bf c)} an effective diffusion coefficient increasing with the
distance between two particles (in the case of relative diffusion).

Mechanism {\bf a)} is perhaps the simplest one and L\'evy flights belong
to this class~\cite{Levy,Metzler}. Some deterministic maps (e.g., the standard 
map for specific values of the control parameter) can produce
super-diffusion according to mechanism {\bf b)}~\cite{standard}. 
An important example of mechanism {\bf c)} is given by the fully
developed turbulence~\cite{Gawedsky}. There, the relative dispersion of 
two particles at distance $R$ is described by a diffusion equation
with an effective diffusion coefficient $D^e \sim R^{4/3}$.
\\
It is natural to wonder about the effects of the super-diffusion
for the front propagation. 
In this letter we discuss mechanism {\bf a)}. 
The evolution equation of $\theta$ has the structure:
\begin{equation}
   \partial_t \theta = 
      \hat{L}_\alpha \theta + f(\theta) / \tau
   \label{eq:evol}
\end{equation}
where $\hat{L}_\alpha$ is a linear operator accounting for the
concentration spreading of test particles evolving according to L\'evy
flights with exponent $\alpha$.  An $\alpha$-L\'evy flight is an
independent increment stochastic process, 
and
the distribution of each increment is a L\'evy-stable distribution
which exhibits a power law tail $P_\alpha(w) \sim |w|^{-(1+\alpha)}$,
with $1<\alpha<2$.  
The moments behave as $\langle |x(t)|^q \rangle \sim
t^{q/\alpha}$ for $q<\alpha$ and $\langle |x(t)|^q \rangle = \infty$
for $q>\alpha$.  So, the role of $\nu$ in Eq.~(\ref{eq:anodif}) is
played by $1/\alpha$ (for $q<\alpha$). Since $1/\alpha > 1/2$, 
one can speak of anomalous diffusion.
Summarizing, we replace the operator 
$-({\bf u} \cdot \nabla) + D_0 \nabla^2$ in the Eq.~(\ref{eq:ard})
by $\hat{L}_\alpha$ substituting an $\alpha$-L\'evy flight
to the Eq.~(\ref{eq:langevin}).
Only for simplicity in the notation and in the numerical computation
\cite{Abel,Torcini} we consider a reacting term which is non zero
only at discrete time step, when $\delta$-form impulses occur:
\begin{equation}
   f(\theta,t) = \sum_{n=-\infty}^{\infty} g(\theta)
                                \delta(t - n\Delta t)\,\Delta t \,\,,
   \label{eq:reactmap}
\end{equation}
and we introduce the reaction map 
$
   G(\theta)=\theta + {\Delta t \over \tau} g(\theta)
$
governing the evolution of an homogeneous field $\theta$
(i.e., without diffusion): $\theta(t+0^+) = G(\theta(t))$. 
The detailed shape of $G(\theta)$ is not
important~\cite{Abel,Torcini}, it is just
necessary to have an unstable fixed point in $\theta = 0$ 
and a stable one in $\theta = 1$. 
In the following we will present the results for the map:
\begin{equation}
G(\theta) = {\theta / [\theta + (1 - \theta)\exp(-\Delta t / \tau)]}\,\,.
\label{eq:reaction}
\end{equation}
i.e., the exact solution 
of the equation $d\theta / dt= \theta(1-\theta)/\tau$.
Noting that, assuming Eq.~(\ref{eq:reactmap}), between $t+0^+$
and $t+\Delta t$ Eq.~(\ref{eq:evol}) reduces to the linear
equation $\partial_t \theta = \hat{L}_\alpha \theta$, and we can
determine the field $\theta(x,t+\Delta t)$ in terms of $\theta(x,t)$:
\begin{eqnarray}
   \theta(x,t+\Delta t) & = & \int_{-\infty}^{+\infty} {\mathrm d}w
        P_{\alpha,\Delta t}(w) \theta(x - w, t+0^+) \nonumber \\
        & = & \int_{-\infty}^{+\infty} {\mathrm d}w 
	P_{\alpha,\Delta t}(w) G(\theta(x-w,t))
   \label{eq:integ}
\end{eqnarray}
where $P_{\alpha,\Delta t}(w)$ is the probability distribution
to have a flight of size $w$ in a time interval $\Delta t$.
Of course $t = n \Delta t$, where $n$ is an integer number.
Let us note that, if one assumes the expression (\ref{eq:reactmap}) for
$f(\theta)$, (\ref{eq:integ}) is an exact relation and not only
an approximation for small $\Delta t$. It is important to stress
that using (\ref{eq:integ}) one can avoid the problem
of the precise definition of $\hat{L}_\alpha$ in terms of fractional
derivative~\cite{derivative,Fogedby}. 
In addition one can study the evolution Eq.~(\ref{eq:integ})
using a generic distribution $P_{\alpha,\Delta t}(w)$ 
which is in the basin of attraction of the $\alpha$-L\'evy stable
distribution.\\
A similar problem has been studied for reaction systems driven
by a L\'evy walk~\cite{Zumofen}.
In \cite{Fedotov} 
reaction-transport processes using
 non gaussian random walk for the diffusion process are studied.
This approach is similar to our one, but the analysis of \cite{Fedotov}
is always for a class of processes which give rise to standard
diffusion. In addition, in the context of disturbance
propagation in chaotic extended systems with long-range coupling,
Torcini and Lepri~\cite{Torcini2} studied a discrete space version
of (\ref{eq:integ}) with a linear shape for $G(\theta)$.
\\
Regarding the numerical simulations we study a 1D grid with 
open boundary conditions and symmetric initial condition 
with $\theta(x,0) \neq 0$ in a small region around $x=0$. 
We show the results obtained with the reaction map given by 
Eq.(\ref{eq:reaction}) and $\tau=1$.
The principal measured observables are the quantity
of inert material, $m(t)$, and the front speed, $v_f$, defined as:
\begin{equation}
   \!\!\!\!\!m(t) = \int_{-\infty}^{+\infty}\!\! {\mathrm d}x \, \theta(x,t) 
   \; \; \; \; \;\;
   v_f = \lim_{t \to \infty} {m(t) \over {2 t}}\,\,.
   \label{eq:material}
\end{equation}
Simple arguments suggest that if $P(w)$ is steep enough, 
e.g., $P(w) \sim \exp(-\beta |w|)$ then the asymptotic 
behaviour of $\theta(x,t)$ is the usual one of the standard 
FKPP equation, i.e., $\theta(x,t) \sim h(x - v_f t)$ where 
$v_f \propto \sqrt{\langle w^2 \rangle}$ and 
$h(z) \sim \exp(-\gamma z)$ for $z \gg 1$. 
\begin{figure}
\epsfig{file=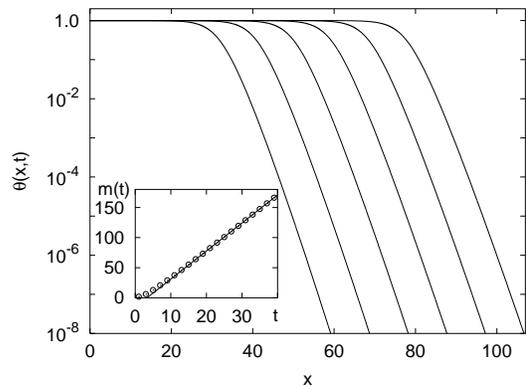,width=0.9\linewidth}
\caption{Front shapes at increasing times in the case of $P(w) \sim
\exp(-\beta|w|)$, with $\beta=1$. The inset shows the quantity
of inert material with its asymptotic linear behaviour
(solid line). We show just one side of the front.}
\label{fig:fig1}
\end{figure}
In this case the inert material increases linearly in time: 
$m(t) = \mbox{const} + 2 v_f t$ (as for the FKPP equation, see Fig.~1).
\\
More interesting is the case of super-diffusive
behaviour with $\alpha$-L\'evy flight ($1 < \alpha < 2$). 
The shape of $P(w)$ used in the numerical computation is 
$P(w)=P_0$ for $|w|<w_0$ and $P(w) = P_n|w|^{-(1+\alpha)}$
for $|w|\geq w_0$. The values of the parameters, $P_n$ and $P_0$, 
are chosen to guarantee continuity in $w_0$ and normalization of $P(w)$. 
Fig.~2 shows that
at large time $\theta$ develops a power law behaviour 
$\theta \sim |x|^{-(\alpha + 1)}$. In addition, the inert material 
invades the fresh one ($\theta \simeq 0$) exponentially
fast. This can be seen by looking at the total quantity of
inert material at time $t$, 
\begin{equation}
   m(t) \sim e^{ct}
   \label{eq:expgro}
\end{equation}
(see the inset of Fig.~2).

\begin{figure}
\epsfig{file=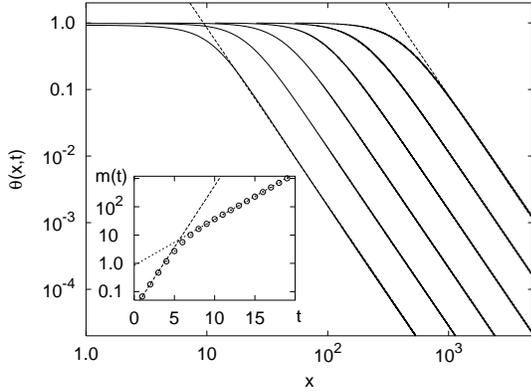,width=0.9\linewidth}
\caption{Front shapes at increasing time in the case of $P(w) \sim
|w|^{-(1+\alpha)}$ with $\alpha ={5 \over 3}$. The dashed lines
indicate the theoretical prediction $x^{-(1+\alpha)}$.
The inset shows the exponential growth of the inert material.
The dotted line is the asymptotic behaviour proportional 
to $a^{t/(1+\alpha)}$ (see the main text) and the dashed one
is the initial rate which grows exponentially as 
$a^{t}$.}
\label{fig:fig2}
\end{figure}
The numerical results illustrated above are supported by an analytical
argument using a linear analysis of the tail of $\theta(x,t)$
(which is expected to be valid for pulled dynamics~\cite{Torcini,Ebert}) 
and the theory of the infinitely divisible distributions~\cite{Gnedenko}. 
For $\theta$ around zero,
$G(\theta)$ has a linear shape
$G(\theta) \simeq a \theta$ with $a>1$.
Plugging this into (\ref{eq:integ}), for $x \gg 1$,
one has
\begin{eqnarray}
   \theta(x,t) & \simeq & a (P * \theta)(x,t-1) \nonumber \\
               & \simeq & a^t (P * P * \cdots * P * \theta)(x,0)
   \label{eq:convol}
\end{eqnarray}
where $*$ indicates the convolution operation. 
It is well known~\cite{Gnedenko} that processes with power law
tail $P(w) \sim |w|^{-(1 + \alpha)}$ with $1 < \alpha < 2$ are in 
the basin of attraction of the $\alpha$-L\'evy-stable distribution
$P_\alpha(w)$. Therefore~(\ref{eq:convol}) yields for
$|x| \gg 1$ and large $t$
\begin{equation}
   \theta(x,t) \sim |x|^{-(1+\alpha)} a^t\,\,,
   \label{eq:asintheta}
\end{equation}
in agreement with the behaviour of Figure 2.
The growth coefficient in Eq.~(\ref{eq:expgro}), $c$,
can be computed with a matching argument. We expect
(and this is confirmed by simulations) there exists an
$\tilde{x}$ value such that $\theta \sim 1$
for $x < \tilde{x}$ and $\theta \sim |x|^{-(1+\alpha)}$
for $x > \tilde{x}$. We can write $m(t) \simeq 2 \tilde{x}$, and
the value of $\tilde{x}$ can be obtained simply matching
the $\theta(x,t)$ in (\ref{eq:asintheta}) with $\theta \simeq 1$,
i.e., $\tilde{x} \sim a^{t/(1+\alpha)}=\exp({\ln a \over 1+\alpha} t)$
and therefore $c=\ln a /(1+\alpha)$, 
in good agreement with numerical 
results (see the inset of Fig.~2).
The transient, when $\max{\theta(x,0)}$ is small enough, 
is dominated by the reaction term, giving an exponential 
growth $m(t) \sim a^{t}$.

Particularly interesting is the behaviour for 
$P(w) \sim |w|^{-(1 + \alpha)}$ with $\alpha > 2$. 
In this case the distribution belongs to the basin
of attraction of the gaussian law
since $\langle w^2 \rangle < +\infty$ (see~\cite{Gnedenko}). 
In fact, although the probability distribution, 
$P(x_t)$, of the sum of independent random variables, 
$x_t = \sum_{j=1}^t w_j$, 
has a power law tail, the core of the distribution behaves
as a gaussian, and the tail is less and less important 
as $t$ grows (see Fig.~3).
Therefore, at first glance, one could expect the same 
features of the standard FKPP equation. 

\begin{figure}
\epsfig{file=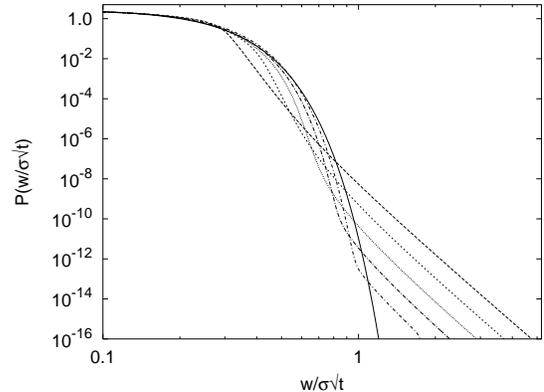,width=0.9\linewidth}
\caption{Rescaled probability density functions, $P(x_t/\sigma \sqrt{t})$, 
of the sum of independent random variables, $x_t = \sum_{j=1}^t w_j$,
in the case of $P(w) \sim |w|^{-(1 + \alpha)}$ with $\alpha = 10$
and $t = 2,4,8,16,32$ from top to bottom; $\sigma$ is the standard 
deviation of $P(w)$.
The solid line is the gaussian asymptotic behaviour. }
\label{fig:fig4}
\end{figure}
But, in reacting systems the presence of this tail
can have an important role. 
In fact for each initial condition, $\theta(x,0)$, localized in a small
region (e.g., around $x=0$), already at the first step, the front
has a shape not steep enough for the usual FKPP propagation 
(i.e., $\theta(x,1) \sim |x|^{-(1+\alpha)}$)~\cite{Ebert}.
Then, because of the reaction, i.e., the instability of $\theta=0$,
the tail of $\theta$ increases exponentially
in time. As consequence of the gaussian core of $P(x_t)$ we expect
that the bulk of $\theta$ behaves in the FKPP way,
but, at large time, the exponential growth of the tail 
has the dominant role.\\
In Fig.~4 it is shown how, the exponential form of 
the front (which moves with a constant velocity) is overcome 
by a power law tail that grows more and more as the time increases,
i.e., the initial FKPP behaviour (constant $v_f$ and 
$\theta(x,t)$ with exponential decay) is replaced by the
exponential increasing of the inert material
and $\theta(x,t)$ with power law tail.
This result is similar to that obtained in~\cite{Paladin} in the context
of the growth of perturbation in CML: even a weak long range
coupling (e.g, a power law with large $\alpha$) has, at large
time, a dramatic effect.\\
Summarizing, we have shown how, in a reaction diffusion system,
replacing a standard diffusion with a process whose
probability distribution has a tail approaching to zero
faster than a power law, one has the same qualitative behaviour
of the usual FKPP system, i.e., exponential decay of the front
and a finite front speed.
\begin{figure}
\epsfig{file=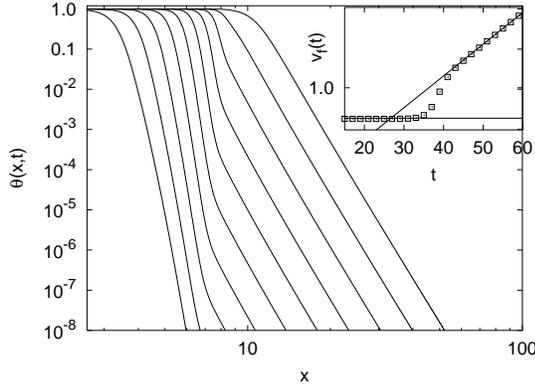,width=0.9\linewidth}
\caption{Front shapes at $t=19, 22, 25 \cdots$ 
where $P(w)$ is the same as Figure 3.  In the inset it is
shown the front speed $v_f(t) = (m(t+1)-m(t))/2$.  The straight lines
indicate the linear propagation regime, $v_f = cost$, and the
exponential propagator regime, $v_f \propto \exp(t \; \ln{a}/(1+\alpha))$.}
\label{fig:fig5}
\end{figure}
 On the contrary, if the process has the 
same tail in the probability distribution of an $\alpha$-L\'evy stable 
process one achieves an exponentially fast propagation of the front,
instead of the linear one. Moreover the tails of the field $\theta$
have a power law behaviour instead of the exponential decay.
As intermediate behaviour we have the case of a stochastic process 
whose increment have a power
law tail but with finite variance. In this case, initially one
has the usual FKPP behaviour, but after a while, one has an exponential
growing of the inert material.
\\
Let us now compare the above described scenario with the case of
super-diffusion induced by strong time correlation of the velocity
field. An example is provided by the standard map
with suitable values of the control parameter $K$~\cite{standard}.
This advection-diffusion system can show super-diffusive
behaviour~\cite{standard} in the limit of $D_0 \to 0$.  However in the
presence of reaction the front speed is finite for each value of
$D_0$, because it is bounded by $U_{max} + v_0$, where $U_{max}$ is
the maximum velocity of the test particle, and $v_0$ is the front
speed in absence of stirring.
Therefore, just the presence of anomalous diffusion does not 
necessarily imply a non-linear front propagation, but also
the details of the transport-diffusion and reaction mechanisms 
are important. For example, in the system (8) 
using a reaction map with $G(\theta) = \theta$ around $\theta=0$
(e.g., ignition)
we have always the usual scenario shown in Fig.~1
also for $1<\alpha<2$. \\
Such a fact is, in our opinion, relevant for the modeling of
realistic reaction systems: one has to mimic precisely the 
mechanism that gives rise to the anomalous diffusion,
beyond the simple power law behaviour of
$\langle x^2(t) \rangle$. 
\\

We thank M.~Cencini, P.~Muratore-Gianneschi and A.~Torcini 
for useful discussions and comments.
A.V. thanks A.Kupiainen for his kind hospitality
at the University of Helsinki during the final write up of this article.
This work has been partially supported by INFM {\it Parallel Computing
Initiative} and MURST ({\it Cofinanziamento Fisica Statistica di
Sistemi Complessi Classici e Quantistici}).  
We acknowledge support from the INFM {\it Center for 
Statistical Mechanics and Complexity} (SMC).


\end{document}